\documentclass[aps,amsmath,amssymb,preprintnumbers,groupedaddress,twocolumn]{revtex4}
\usepackage{graphicx}
\newcommand{\nc}{\newcommand}
\nc{\beq}{\begin{equation}} \nc{\eeq}{\end{equation}} \nc{\bea}{\begin{eqnarray}}
\nc{\eea}{\end{eqnarray}}

\def\gsim{\mathrel{\rlap{\lower4pt\hbox{\hskip1pt$\sim$}}
    \raise1pt\hbox{$>$}}}       %greater than or approx. symbol

\def\K3{{\bf K3}}

\def\ov{\overline}

\def\n2d{\cN_{V^*}^{\otimes 2}}

\def\cN{{\mathcal N}}

\def\ch{\mbox{ch}}

\def\sym{{\rm Sym}}
\def\anti{{\rm Anti}}

\begin{document}

\preprint{UPR-1188-T}
\title{Hierarchies from D-brane instantons in globally defined Calabi-Yau Orientifolds}

\author{Mirjam Cveti{\v c}} %\email{cvetic@cvetic.hep.upenn.edu}
\author{Timo Weigand}%\email{timo@physics.upenn.edu}
\affiliation{Department of Physics and Astronomy, University of Pennsylvania, Philadelphia, USA}

\begin{abstract}
\noindent We construct the first globally consistent semi-realistic  Type I string vacua on an
elliptically fibered manifold where the zero modes of the Euclidean D1-instanton sector allow for the
generation of non-perturbative Majorana masses of an intermediate scale. In another class of global
models, a D1-brane instanton can generate a Polonyi-type superpotential breaking supersymmetry at an
exponentially suppressed scale.

\end{abstract}

%\pacs{}

\maketitle

\date{today}

\bigskip

\section{Introduction}

The quest for a natural generation of hierarchies is a classic theme in particle physics. Prominent
examples of  hierarchical mass scales are, amongst others, intermediate scale Majorana masses for
right-handed neutrinos or the  $\mu$-term  and supersymmetry breaking scale in the minimal
supersymmetric Standard Model. In the past year important conceptual progress has been made
 towards realizing these hierarchies due to new D-brane instantons  \cite{Blumenhagen:2006xt,Haack:2006cy,Ibanez:2006da,Florea:2006si}  of genuinely string theoretic origin with apparently no field theory analogs.  In type II  string compactifications with D-branes the mentioned couplings are typically   forbidden perturbatively, but  under specific conditions they can be generated by D-brane instantons whose strength is exponentially suppressed by the classical instanton action with respect to the string scale.
This transforms the mass hierarchies into logarithmic ones. Further work in this direction includes \cite{Abel:2006yk,Akerblom:2006hx,Bianchi:2007fx,Cvetic:2007ku,Ibanez:2007rs,Argurio:2007vqa,Bianchi:2007wy,Akerblom:2007uc,Blumenhagen:2007zk,Billo:2007py}.

While local realisations of this  mechanism  have been found, see, e.g., 
\cite{Cvetic:2007ku},  an important outstanding  challenge lies in the 
construction of globally
 consistent chiral string vacua which  exemplify  
explicitly the non-perturbative
origin of the respective hierarchies. %For efforts within 
%semi-realistic Gepner models see \cite{Ibanez:2007rs}.
%The main difficulty  is due to the tension between supersymmetry and global consistency conditions in
%the D-brane sector and specific constraints on the allowed fermionic zero modes in the instanton
%background; typically, additional D-branes, required by global consistency conditions, introduce
%extra fermionic zero modes that ``annihilate'' the instanton effect. Furthermore,  the classical
%instanton action depends on the closed string sector moduli which are constrained by the D-brane
%sector supersymmetry conditions. Thus, even before tackling full moduli stabilisation,  the desired
%range of  exponentially suppressed couplings has to be compatible with the supersymmetry constraints.
The main difficulty  is due to tadpole cancellation conditions corresponding to Gauss law: the total charge of the D-branes on the compact internal manifold
has to vanish. Typically this requires additional D-branes in a hidden sector. These can introduce extra fermionic instanton zero modes in form of open strings between the D-branes and the instanton that ``annihilate'' the instanton effect. Furthermore,  the classical
instanton action given by the volume of the instanton depends on the shape or size of the manifold, i.e. the moduli fields; these are constrained by the D-brane
sector supersymmetry conditions. Thus, even before tackling full moduli stabilisation,  the desired
range of  exponentially suppressed couplings has to be compatible with the supersymmetry constraints.

In this letter we provide the first classes of semi-realistic globally consistent string vacua with
D-branes where some hierarchical couplings can be realised by D-brane instantons. We choose the
framework of Type I string  theory 
%compactified on an elliptically fibered Calabi-Yau space
with
magnetized D9-branes compactified on a specific Calabi-Yau threefold $X$.
%associated
% with holomorphic gauge bundles amenable to algebraic geometry techniques.
 The Euclidean D1-branes (E1-instantons) wrap isolated holomorphic curves on $X$.
 For a suitable choice of compactification manifold the charged zero mode
 structure of the instanton can be engineered to allow for the existence of
 specific non-perturbative couplings.  Our focus is to demonstrate that it is indeed possible to
satisfy all global consistency conditions and maintain the required instanton zero mode structure
while the desired suppression scale is compatible with the D-brane sector supersymmetry constraints.
To this aim we have not attempted to build a fully-realistic Standard Model sector but content
ourselves with SU(5) Grand Unified (GUT) toy models with four chiral families.
 While the models  have phenomenological drawbacks, such as a large number of Higgs pairs
 and chiral exotics,  they
demonstrate  for the first time an explicit global realisation of the
 described instanton physics along with a sufficiently complex particle physics sector.
 The present framework can be further refined to build ever more
 realistic string vacua with D-instanton effects.

Our first class of examples realizes intermediate scale Majorana masses. In the  second class of
examples instantons generate a Polonyi-type superpotential in a hidden sector which breaks
supersymmetry at an exponentially suppressed scale. These techniques can readily be applied to
realize other instanton effects such as the generation of $\mu$-terms
\cite{Blumenhagen:2006xt,Ibanez:2006da} or certain GUT Yukawa couplings \cite{Blumenhagen:2007zk}.

\section{E1-instantons in Type I vacua}

\noindent We consider Type I compactifications on an internal Calabi-Yau threefold $X$. The gauge
sector is defined in terms of stacks of $M_a=n_a \times N_a$ spacetime-filling D9-branes wrapping the
whole of $X$ (and their orientifold images), where $\sum_a n_a N_a = 16$.  These D9-branes
can carry rank $n_a$ holomorphic vector bundles $V_a$ whose structure group $U(n_a)$ breaks the
original gauge group $U(M_a)$ associated with the coincident D9-branes to the commutant $U(N_a)$
\cite{Blumenhagen:2005zg, Blumenhagen:2005zh}. Stacks of $N_i$ D5-branes wrapping the holomorphic
curve $\Gamma_i$ on $X$ are described by the sheaf $i_* {\cal O}|_{\Gamma_i}$ supported
on $\Gamma_i$ and carry gauge group $Sp(2N_i)$.

The massless open string spectrum is encoded in various cohomology (or rather extension
\cite{Katz:2002gh}) groups associated with the respective bundles on the branes as summarized in
table \ref{Tchiral1}, with the goups of degree 2 (1) counting (anti-)chiral  superfields. For details see \cite{Blumenhagen:2005zh}.
\begin{table}[htb!] 
\renewcommand{\arraystretch}{1.5}
\begin{center}
\begin{tabular}{|c||c|}
\hline \hline
reps. & $\prod_{a} SU(N_a)\times U(1)_a \times \prod_{i} SP(2N_i)$   \\
\hline \hline
$(\sym_{U(N_a)})_{2(a)}$ & $H^*(X,\bigwedge^2 V_a)$  \\
$(\anti_{U(N_a)})_{2(a)}$ & $H^*(X,\bigotimes^2_s  V_a)$  \\
\hline
$(N_a,N_b)_{1(a),1(b)}$ & $H^*(X,V_a \otimes V_b)$ \\
$(\ov N_a,N_b)_{-1(a),1(b)} $ &  $H^*(X,V_a^\vee \otimes V_b)$ \\
\hline
%%%%%%%%%%%%%%%
$(N_a,2N_i)_{1(a)}$ & $H^*(\Gamma_i,V_a^{\vee}|_{\Gamma_i} \otimes K^{1/2}_{\Gamma_i})$ \\
\hline
\end{tabular}
\caption{\small Massless spectrum of Type I compactifications. } \label{Tchiral1}
\end{center}
\end{table}
Cancellation of D5-tadpoles requires $\sum_a N_a \ch_2(V_a) - \sum_i N_i \gamma_i = - c_2(TX)$, and
absence of global anomalies is ensured %by the K-theory constraint 
if $\sum_a N_a c_1(V_a) \in H^2(X,2{\mathbb Z})$.

The vector bundles must be supersymmetric with respect to the O9-plane. For stable holomorphic
bundles this amounts to satisfying the D-flatness condition inside the K\"ahler cone, $\int_X {1
\over 2}     J\wedge J \wedge c_1(V_a)  -
 \ell_s^4 \,
\left( {\rm ch}_3 (V_a)+\frac{1}{24}\, c_1(V_a)\,  c_2(T)\right)=0$, which can be read as a
constraint on the K\"ahler form $J$. Finally the real part of the gauge kinetic function $Re(f_a)=
{\widetilde f}_a/(2 \pi g_s \ell_s^6)$ has to be positive. Here \bea \label{Gauged} {\widetilde f}_a=
{n\over 3!}\,
 \int_X J \wedge J \wedge J -
 \ell_s^4\,  \int_X J \wedge
\left( {\rm ch}_2 (V_a)+\frac{n_a}{24}\, c_2(T)\right). \nonumber \eea

The superpotential of the four-dimensional ${\cal N}=1$ supersymmetric effective action receives
non-perturbative corrections due to E1-instantons on a  holomorphic curve $C$ \cite{Witten:1999eg}.
For the first related work in this context see \cite{Blumenhagen:2006xt,Bianchi:2007fx}. The universal part of the E1-instanton measure is given by $d^4x \, d^2
\theta$ as is necessary for contributions to the superpotential. To avoid complications due to
additional  instanton bosonic and fermionic deformation zero modes, we focus on E1-branes
wrapping rigid ${\mathbb P}^1$s, where these modes are absent.

In the presence of D9/D5-branes additional fermionic zero modes arise which are charged under the
gauge group on the branes \cite{Blumenhagen:2006xt,Ibanez:2006da,Florea:2006si}. Consider first the
zero modes in the D9-E1 sector. In analogy to the D9-D5 spectrum they should be described by the
cohomolgy groups $H^*(C, V^{\vee}_a|_C \otimes K_C^{1/2})$. The crucial difference as compared to the
spectrum between pairs of D-branes is that only states corresponding to \emph{chiral} operators (from
a worldsheet perspective) are present  \cite{Cvetic:2007ku} and that the four-dimensional
polarisation of the fermionic zero modes is given by a single Grassmann number. As a result, only the
even degree cohomology group $H^0(C, V^{\vee}_a|_C \otimes K_C^{1/2})$ corresponds to actual
fermionic zero modes in the representation $({N}_a,1_E)$, while the would-be modes classified by
$H^1(C, V^{\vee}_a|_C \otimes K_C^{1/2})$ are not realised physically. Zero modes in the conjugate
representation $(\ov N_a,1_E)$ are associated with $H^0(C, V_a|_C \otimes K_C^{1/2}) = H^1(C,
V^{\vee}_a|_C \otimes K_C^{1/2})^*$. The last equality follows from Serre duality on $C$. All this is
summarized in table \ref{lambda_cohom}.
\begin{table}[htb!]
\renewcommand{\arraystretch}{1.5}
\begin{center}
\begin{tabular}{|c|c|c|}
\hline \hline
state & rep & cohomology    \\
\hline \hline
$\lambda_a$  & $(N_a,1_E)$   & $H^0({\mathbb P}^1, V_a^{\vee}(-1)|_{{\mathbb P}^1})$  \\
$\ov\lambda_a$  & $(\ov N_a,1_E)$   & $H^1({\mathbb P}^1, V_a^{\vee}(-1)|_{{\mathbb P}^1})^*$ \\
\hline \hline
\end{tabular}
\caption{\small Fermionic zero modes in D9-E1 sector. } \label{lambda_cohom}
\end{center}
\end{table}
For $C$ topologically just a ${\mathbb P}^1$, $K_{{\mathbb P}^1} = {\cal O}(-2)$, and $
V_a|_{{\mathbb P}^1} \otimes {\cal O}(-1) = V_a(-1)|_{{{\mathbb P}^1}} = \bigoplus_{i=1}^{n_a} {\cal
O}(k_i -1)$ with $\sum_{i=1}^{n_a} k_i = \int_{{\mathbb P}^1} c_1(V_a)$. In general, the splitting
type of the bundle $V_a$ varies over the bundle moduli space. However, for the special case of a line
bundle $L_a$, it can directly be read off from  $x_a = \int_{{\mathbb P}^1} c_1(L_a)$ as  $
L_a(-1)|_{{{\mathbb P}^1}} = {\cal O}(x_a-1)$. The charged zero modes are then  determined  by Bott's
theorem, $h^0({\mathbb P}^1,{\cal O}(k)) = \theta(k) (k+1), h^1({\mathbb P}^1,{\cal O}(k)) =
\theta(-k) (-k-1),$ with $\theta(k)=1$ for $k \geq 0$ and else zero.

Extra fermionic zero modes from the D5-E1 sector are
 counted by the extension groups $Ext_X(j_* {\cal
O}|_{\Gamma_i},i_* {\cal O}|_C)$. These groups vanish when $\Gamma_i$ and $C$ do not intersect.

If all  $\lambda$ modes can be absorbed consistently
\cite{Blumenhagen:2006xt,Ibanez:2006da,Florea:2006si}, the E1-instanton yields contributions to the
superpotential of the schematic form \bea \label{super} W=  M_s^{3-k} \prod_{i=1}^k \Phi_i e^{-
\frac{2 \pi}{g_s} {{\mathcal Vol} \over \ell_s^2} } =  M_s^{3-k} \prod_i \Phi_i  e^{-  \frac{2 \pi \,
\ell_s^4 }{\alpha_{a}} \frac{{\mathcal Vol}}{{\widetilde f_a}} }. \eea Here we traded $g_s$ in for
the gauge coupling on a reference brane $D9_a$.
%, and $M_s^2 = 2 \pi m_G^2 \alpha_{GUT}^2 {\widetilde f_a}/{\ell_s^6}$ with $m_G = 2.4 \times 10^{18}$ GeV.
The  scale of the non-perturbative term is thus
controlled by the ratio of the instanton volume ${\mathcal Vol}= \int_{{\mathbb P}^1} J$ to the gauge
kinetic function ${{\widetilde f_a}}$. This ratio
is a function of the K\"ahler moduli which are in general only partially constrained by the
D-flatness conditions. In (\ref{super}) we also  suppressed the possible dependence on the complex
structure moduli through the one-loop Pfaffian \cite{Blumenhagen:2006xt,Akerblom:2006hx}. While the interaction
scale is fully determined only  for completely stabilised moduli, we have to content ourselves, in
the presence of unfixed moduli, with showing that the hierarchies of the instanton effects are
compatible with the D-term conditions for the D9-branes.

\section{E1-instantons on elliptic CY3}

\noindent We now present a class of  globally defined supersymmetric models exhibiting such instanton
effects. We choose a Calabi-Yau threefold $X$ which is elliptically fibered (with fibration $\pi$) over the del Pezzo
surface $dP_r$ for $r=4$. For detailed information on this 
type of geometries see, e.g., \cite{Donagi:1999gc}. For concrete 
algebraic geometry formulae relevant in our context see 
\cite{Blumenhagen:2005zg}.  The K\"ahler form $J$ enjoys the expansion 
$J/\ell_s^2= r_{\sigma} \sigma + r_l \pi^* l +
\sum_{i=1}^4 r_i \pi^* E_i$ (c.f., Appendix A of \cite{Blumenhagen:2005zg}) in 
terms of the fibre class
$\sigma$, the hyperplane class $l$ as well 
the class $E_i$ of the $4$ ${\mathbb P}^1$ inside $dP_4$
obtained as the blow-up of certain singularities in ${\mathbb C}{\mathbb P}^2$.

This type of geometry is particularly suitable for constructing models exhibiting the desired
non-perturbative corrections: each bundle $V_a$ can
\emph{locally} be engineered in such a way as to lead to the correct number of charged zero modes
with the instanton. This setup can then be promoted to a \emph{globally} 
defined model if the tadpole
cancellation condition can be satisfied without introducing 
additional zero modes between
the instanton and the filler D5-branes. 
To demonstrate the existence of globally consistent models in this context it suffices to consider
branes carrying line bundles $L_a$ with first Chern class $c_1(L_a)= q_a \sigma + \pi^*\zeta_a$ with $\zeta_a \in H^2(dP_4, {\mathbb
Z})$. For the concrete expressions of all higher Chern characters and the Euler characteristic
$\chi(L_a)$ in terms of $c_1(L_a)$  see sections 3.3 and 3.4 
of \cite{Blumenhagen:2005zg}, 
while the full cohomology groups $H^*(X, L_a)$ can be computed  from 
appendix B in \cite{Blumenhagen:2006wj}.

We consider instantons wrapping  certain rigid non-horizontal ${\mathbb P}^1$s 
in the $dP_9$ surface $\pi^*E_4$, which is obtained as the pullback of the ${\mathbb P}^1$ $E_4$ in the base. 
Specifically, the ${\mathbb P}^1$s are taken not to intersect the base of $dP_9$.
%This choice ensures 
%that for the above type of line bundles one can always take each of the filler D5-branes to wrap a purely horizontal or vertical curve $\Gamma_i$ that
%does not hit the 
%instanton, thus introducing no extra zero-modes. 
It turns out that one can always take the filler D5-branes to wrap curves $\Gamma_i$ that do not hit at least $\emph{some}$ of the ${\mathbb P}^1$s in that class, thus introducing no extra zero modes for this subset of instantons.

\subsection{Majorana masses}

\noindent In models with instanton-generated Majorana masses a particular set of magnetized D9-branes
engineers the Standard Model (or a GUT version of it), while the right-handed neutrinos $N_R^c$ arise
as the bi-fundamental matter between a pair of $U(1)$ stacks with gauge groups $U(1)_b$ and $U(1)_c$.
 For $N_R^c$ transforming as, say, $(-1_b, 1_c)$, Majorana masses can be generated in the presence of precisely 2 charged fermionic instanton zero modes of type $ \lambda_b$ and $\ov \lambda_c$ \cite{Blumenhagen:2006xt,Ibanez:2006da}.
This guarantees that the coupling $\lambda_b N_R^c \ov\lambda_c$ in the instanton moduli action is
allowed by gauge invariance. Integration over 
two copies of $\lambda$-modes then generates a mass term
for $N_R^c$. The actual presence of the zero 
mode couplings in the moduli action is due to classical overlap integrals 
and we have checked that they indeed exist. Further details as well as the resulting family structure will be 
presented elsewhere.

For realisations of the $U(1)_b$ and $U(1)_c$ stack in terms of single D9-branes carrying line
bundles $L_b$ and $L_c$, the zero mode constraints translate into $h^i({\mathbb P}^1,L_b^{\vee}(-1)|_{E_4}) =
(2,0) = h^i({\mathbb P}^1,L_c(-1)|_{E_4})$. For the described subclass of ${\mathbb P}^1$s in  $\pi^*E_4$ not intersecting the horizontal $E_4$ in $dP_4$, this can be shown to correspond to
%Due to Bott's theorem  this is equivalent to
$\zeta_b \cdot E_4 = -2 = - \zeta_c \cdot E_4$.
%Absence of any further charged zero 
%modes %between the Standard Model sector and the instanton,
%between filler D5-branes and the instanton is guaranteed since the instanton wraps a non-horizontal curve.
%by the choice of the 
%instanton as wrapping a ${\mathbb P}^1$ curve  in the del Pezzo surface 
%$dP_9$ obtained as an elliptic fibration over $E_4$.
Since our primary aim is to find global embeddings of the instanton sector, we content ourselves with
the realisation of a GUT toy version of the Standard Model sector, namely to engineer it as a U(5)
theory from $N_a = 5$ D9-branes with line bundle $L_a$.
%It turns out that the above constraints can easily be met in the present context.
In table \ref{model_Maj} we give a representative example of an SU(5) model of the type described.
All D5-brane tadpoles are cancelled by including also stacks of $N_i$ unmagnetized D5-branes on
curves $\Gamma_i$ with total D5-brane charge $\sum N_i \gamma_i = 41 F + \sigma \cdot \pi^*(16 l -12 E_1)$, with $F$ the fibre class. Furthermore,
12 unmagnetized D9-branes are required to cancel the D9-brane tadpoles. One can check that the
D-flatness conditions allow for solutions inside the K\"ahler cone, e.g.,
for $r_{\sigma}=0.97, r_l =
10.49, r_1=-7.27$ and $r_2=r_3=r_4=-1.00$.
\begin{table}[htb!]
\renewcommand{\arraystretch}{1.5}
\begin{center}
\begin{tabular}{|c|c|c|}
\hline \hline
Bundle & N & $c_1(L)= q \sigma + \pi^*(\zeta)$    \\
\hline \hline $L_a$ & 5 & $\pi^*(-2 E_3)$ \\\hline $L_b$ & 1 & $2 \sigma + \pi^*(- 2l -2 E_1 + 3E_2 + 2E_3+ 2 E_4)$
\\\hline
$L_c$ & 1 & $-2 \sigma + \pi^*(2l - E_2 - 2 E_3 - 2 E_4)$    \\
\hline
%%%%%%%%%%%%%%%
\hline
\end{tabular}
\caption{\small A $ U(5) \times U(1) \times U(1)$ model with Majorana masses.  } \label{model_Maj}
\end{center}
\end{table}
The spectrum can  be computed  from table \ref{Tchiral1} and the formulae of
\cite{Blumenhagen:2005zg,Blumenhagen:2005zh}. It contains four chiral families of $\bf {\ov {10}}$
counted by $H^*(X, (L_a^{\vee})^2)$ together with additional ${\bf 5}$ and ${\bf \ov 5}$ from the
$a-b$ and $a-c$ sector as well as from the filler D5- and D9-branes. Only part of them can be
interpreted as matter ${\bf 5}$ and Higgs pairs once Yukawa couplings are taken into account.
%What is more of interest for us is the appearance of
Four chiral generations transforming as $(-1_b,1_c)$ in the spectrum
 %and counted by $\chi(X, L_b^{\vee} \otimes L_c)=4$.
 correspond to  right-handed neutrinos $N_R^c$. The zero mode structure of 
the  $E1$ instanton ensures that a superpotential 
of the form $W =x \,  M_s  
\exp({-  \frac{2 \pi \,
\ell_s^4 }{\alpha_{GUT}} \frac{{\mathcal Vol}}{{\widetilde f_a}} })\,  
N_R^c \,  N_R^c$ can be generated,
where $x$ is an ${\cal O}(1)$ factor from the exact computation. 
The above value of the K\"ahler
moduli corresponds to ${\widetilde f_a} = 8.15 \, \ell_s^6$ and an instanton volume 
${\mathcal Vol}=-r_4=1$. Then for $\alpha_{GUT}\sim 0.04$, corresponding to  $M_s \sim 10^{18}\, $GeV, the 
Majorana mass is  ${\cal O}(10^{10}) \, $GeV. Thus, we 
demonstrated that 
in a global setup the 
exponential suppression can indeed be engineered
at the required intermediate scale.  %While this is true for each single  instanton that wraps a ${\mathbb P}^1$ curve in the  
%described class,
%, and satisfies the zero mode constraints, 
The full answer involves summing up all instanton contributions 
associated  with all
suitable  curves, possibly along the 
lines of \cite{Donagi:1996yf}.  Note that potentially dangerous vanishing theorems can be bypassed by suitably distributing the D5-branes such that they introduce additional zero-modes on {\emph {some}} of the rigid ${\mathbb P}^1$ of the instanton class. This will be detailed elsewhere.

\subsection{Polonyi model}

\noindent A fascinating application of stringy instanton effects lies in the context of realising
supersymmetry breaking at a naturally suppressed scale. In the probably simplest such scenario
supersymmetry is broken by a Polonyi-type superpotential of the form $W= c \, \Phi$ for a superfield
$\Phi$. The possible generation of such tadpoles for open string fields by D-brane instantons was
first pointed out in \cite{Blumenhagen:2006xt} and discussed recently in local setups in
\cite{Aharony:2007db,Aganagic:2007py}. As a  variant of the configuration  leading to Majorana
masses, the Polonyi field $\Phi$ arises in the bi-fundamental sector of two massive U(1) gauge groups
$U(1)_b$ and $U(1)_c$ of charge, say, $(-1_b, 1_c)$, while the linear superpotential is generated by
an E1-instanton with the two zero modes $\lambda_b$ and $\ov \lambda_c$. In table \ref{model_Pol}  we
present a global embedding of this scenario into an $SU(5)$ GUT-type vacuum with 4 chiral generations
of {$\bf \ov {10}$} and, as before, further ${\bf 5}$ and ${\bf \ov 5}$.
\begin{table}[htb!]
\renewcommand{\arraystretch}{1.5}
\begin{center}
\begin{tabular}{|c|c|c|}
\hline \hline
Bundle & N & $c_1(L)= q \sigma + \pi^*(\zeta)$    \\
\hline \hline $L_a$ & 5 & $\pi^*( -2 E_1 + 2 E_2 - 2 E_3) $ \\\hline $L_b$ & 1 & $ -2\sigma +\pi^*(l + E_1 - 2 E_2 +
E_3 + E_4)$ \\\hline $L_c$   & 1 & $-4 \sigma + \pi^*(l-E_1  - E_3 - E_4)$    \\\hline
%%%%%%%%%%%%%%%
\hline
\end{tabular}
\caption{\small A $ U(5) \times U(1) \times U(1)$ Polonyi-type model.  } \label{model_Pol}
\end{center}
\end{table}

The tadpole constraints require the introduction of five-brane stacks of total class $ \sum_i N_i
\gamma_i = 28 F + \sigma \cdot \pi^*( 2\, E_2)$. The bundle configuration is D-flat inside the K\"ahler cone
e.g., for $r_{\sigma}=1.59$, $r_l= 12.67$, $r_1=-2.11$, $r_2=-5.00$, $r_3= -4.00$, $r_4 = 
-6.00$, for which $\widetilde
f_a= 38.78 \,  \ell_s^6$. The  zero mode structure of an 
instanton wrapping a non-horizontal ${\mathbb P}^1$  curve of the type described in $\pi^*E_4$ 
%the $dP_9$  
%lliptically fibered over  $E_3$ or $E_4$ is precisely of the
 %described 
allows for 
the generation of a Polonyi-type term  for $\Phi$ of
the form $W=  x M_s^2 \, \exp ({-S_{E1}})\, 
 \Phi$. For K\"ahler moduli of 
the above  D-flat value, the instanton volume is ${\mathcal Vol}=6$, and the
Polonyi-term  is suppressed by $\exp(-S_{E1})=\exp(-  \frac{2 \pi \, 
\ell_s^4 }{\alpha_{GUT}}
\frac{{\mathcal Vol}}{{\widetilde f_a}}) \simeq 2.8 \times 10^{-11}$. Its F-term breaks supersymmetry
dynamically at the scale $F_\Phi\sim 10^{-11}\, M_{s}^2$. 
The particular virtue of the
 D-brane instanton lies in engineering this hierarchical
scale naturally from the
 string theoretical point of view \cite{Aharony:2007db,Aganagic:2007py}.
For gravity mediation this can lead to soft supersymmetry breaking 
masses %$m_{soft} \sim (1 - 100)\,  {\rm TeV}$
in the TeV range for $M_{s} \sim (10^{17} - 10^{18})\, $GeV  and 
a suitable matter K\"ahler 
potential, e.g.  as in  \cite{Kallosh:2006dv}. 
%of the form  ${\cal O} (1) \, \Phi_a
%\Phi_a^*$ (both assumptions are 
%justified for closed sector moduli stabilised away from the
%boundaries of moduli space) \cite{Kallosh:2006dv}. 
 Gauge mediated supersymmetry breaking can  also occur  at the
loop-level due to perturbative superpotential couplings of the type $\Phi\,  {\bf {\bar 5}_M\, 5_M}$
where ${\bf {\bar 5}_M}$, ${\bf 5_M}$ represent messenger fields. 
%In our context these field
%candidates are chiral, and in principle 
%it is hard to distinguish between  the role of messenger
%fields and the GUT matter and Higgs fields.  
These % gauge mediated supersymmetry breaking 
effects
% $F\sim \textstyle{\frac{\alpha_{GUT}}{4\pi}} 10^{-16} M_{s}^2$  have
have an extra  suppression $\textstyle{\frac{\alpha_{GUT}}{4\pi}}$ factor. %due to  loop-effects.

We conclude with a few remarks related to the interplay of closed sector moduli stabilization and the
D-instanton induced Polonyi terms. The closed string sector K\"ahler moduli can in principle be
stabilized due to the strong gauge dynamics associated with the additional  $Sp(2N_i)$  gauge group factors. Their
large negative beta  functions induce gaugino condensation and typically result in
``race-track''-type K\"ahler moduli stabilisation, e.g.  along the lines 
of \cite{Cvetic:2003yd}. The  one-loop threshold corrections to the gauge
kinetic function  can lead also to  complex structure moduli  stabilisation, possibly in connection
with fluxes.
These global constructions  therefore provide an intriguing  
framework where strong gauge dynamics
can yield a vacuum solution with stabilized closed sector moduli. 
For unbroken supersymmetry in the
closed string sector the D-instanton induced Polonyi term provides an  
exponentially suppressed
SUSY breaking and possibly even ``uplifting 
mechanism'' \cite{Dudas:2006gr,Abe:2006xp,Kallosh:2006dv}.

\emph{Acknowledgements:} We thank R.Blumenhagen, K. Bobkov,  S. Kachru, J. 
Kumar, R. Richter, E. Sharpe and E. Silverstein for 
discussions, and  especially R. Donagi and  T. Pantev for discussions on elliptic fibrations.
%We also thank the
This research was supported by
DOE Grant EY-76-02-3071. %, and Fay R. and Eugene L. Langberg Chair funds.

%\newpage

%%%%%%%%%%%%%%%%%%%%%%%%%%%%%%%%%%%%%%%%%%%%%%%%%%%%%%%%%%%%%%%%%
%%%
%%%                     BIBLIOGRAPHY
%%%
%%%%%%%%%%%%%%%%%%%%%%%%%%%%%%%%%%%%%%%%%%%%%%%%%%%%%%%%%%%%%%%%%

%\newpage
%\vskip .75 in


\baselineskip=1.6pt
\begin{thebibliography}{99}



%\cite{Blumenhagen:2006xt}
\bibitem{Blumenhagen:2006xt}
  R.~Blumenhagen, M.~Cveti{\v c} and T.~Weigand,
  %``Spacetime instanton corrections in 4D string vacua - the seesaw mechanism
  %for D-brane models,''
  Nucl.\ Phys.\  B {\bf 771} (2007) 113
  [arXiv:hep-th/0609191].
  %%CITATION = NUPHA,B771,113;%%ç

  %\cite{Haack:2006cy}
\bibitem{Haack:2006cy}
  M.~Haack et al., %, D.~Krefl, D.~L\"ust, A.~Van Proeyen and M.~Zagermann,
  %``Gaugino condensates and D-terms from D7-branes,''
  JHEP {\bf 0701}, 078 (2007)
  [arXiv:hep-th/0609211].
  %%CITATION = JHEPA,0701,078;%%


%\cite{Ibanez:2006da}
\bibitem{Ibanez:2006da}
  L.~E.~Ib{\'a}{\~n}ez and A.~M.~Uranga,
  %``Neutrino Majorana masses from string theory instanton effects,''
  JHEP {\bf 0703} (2007) 052
  [arXiv:hep-th/0609213].
  %%CITATION = JHEPA,0703,052;%%

%\cite{Florea:2006si}
\bibitem{Florea:2006si}
  B.~Florea, S.~Kachru, J.~McGreevy and N.~Saulina,
  %``Stringy instantons and quiver gauge theories,''
  JHEP {\bf 0705} (2007) 024
  [arXiv:hep-th/0610003].
  %%CITATION = JHEPA,0705,024;%%








%\cite{Abel:2006yk}
\bibitem{Abel:2006yk}
  S.~A.~Abel and M.~D.~Goodsell,
  %``Realistic Yukawa couplings through instantons in intersecting brane
  %worlds,''
  JHEP {\bf 0710} (2007) 034
  [arXiv:hep-th/0612110].
  %%CITATION = JHEPA,0710,034;%%

%\cite{Akerblom:2006hx}
\bibitem{Akerblom:2006hx}
  N.~Akerblom et al., % R.~Blumenhagen, D.~Lust, E.~Plauschinn and M.~Schmidt-Sommerfeld,
  %``Non-perturbative SQCD Superpotentials from String Instantons,''
  JHEP {\bf 0704} (2007) 076
  [arXiv:hep-th/0612132].
  %%CITATION = JHEPA,0704,076;%%



%\cite{Bianchi:2007fx}
\bibitem{Bianchi:2007fx}
  M.~Bianchi and E.~Kiritsis,
  %``Non-perturbative and Flux superpotentials for Type I strings on the Z_3
  %orbifold,''
  Nucl.\ Phys.\  B {\bf 782}, 26 (2007)
  [arXiv:hep-th/0702015].
  %%CITATION = NUPHA,B782,26;%%



%\cite{Cvetic:2007ku}
\bibitem{Cvetic:2007ku}
  M.~Cveti\v c, R.~Richter and T.~Weigand,
  %``Computation of D-brane instanton induced superpotential couplings -
  %Majorana masses from string theory,''
  Phys.\ Rev.\  D {\bf 76}, 086002 (2007)
  [arXiv:hep-th/0703028].

%\cite{Ibanez:2007rs}
\bibitem{Ibanez:2007rs}
  L.~E.~Ib{\'a}{\~n}ez, A.~N.~Schellekens and A.~M.~Uranga,
  %``Instanton Induced Neutrino Majorana Masses in CFT Orientifolds with
  %MSSM-like spectra,''
  JHEP {\bf 0706}, 011 (2007)
  [arXiv:0704.1079 [hep-th]].
  %%CITATION = JHEPA,0706,011;%%



%\cite{Argurio:2007vqa}
\bibitem{Argurio:2007vqa}
  R.~Argurio et al., %, M.~Bertolini, G.~Ferretti, A.~Lerda and C.~Petersson,
  %``Stringy Instantons at Orbifold Singularities,''
  JHEP {\bf 0706} (2007) 067
  [arXiv:0704.0262 [hep-th]].
  %%CITATION = JHEPA,0706,067;%%

%\cite{Bianchi:2007wy}
\bibitem{Bianchi:2007wy}
  M.~Bianchi, F.~Fucito and J.~F.~Morales,
  %``D-brane Instantons on the T^6/Z_3 orientifold,''
  JHEP {\bf 0707} (2007) 038
  [arXiv:0704.0784 [hep-th]].
  %%CITATION = JHEPA,0707,038;%%






%\cite{Akerblom:2007uc}
\bibitem{Akerblom:2007uc}
  N.~Akerblom et al., %   , R.~Blumenhagen, D.~Lust and M.~Schmidt-Sommerfeld,
  %``Instantons and Holomorphic Couplings in Intersecting D-brane Models,''
  JHEP {\bf 0708} (2007) 044
  [arXiv:0705.2366 [hep-th]].
  %%CITATION = JHEPA,0708,044;%%

%\cite{Blumenhagen:2007zk}
\bibitem{Blumenhagen:2007zk}
  R.~Blumenhagen et al., %, M.~Cveti{\v c}, D.~L\"ust, R.~Richter and T.~Weigand,
  %``Non-perturbative Yukawa Couplings from String Instantons,''
  arXiv:0707.1871 [hep-th].
  %%CITATION = ARXIV:0707.1871;%%

%\cite{Billo:2007py}
\bibitem{Billo:2007py}
  M.~Billo et. al., %M.~Frau, I.~Pesando, P.~Di Vecchia, A.~Lerda and R.~Marotta,
  %``Instanton effects in N=1 brane models and the Kahler metric of twisted
  %matter,''
  JHEP {\bf 0712} (2007) 051
  [arXiv:0709.0245 [hep-th]].
  %%CITATION = JHEPA,0712,051;%%








%\cite{Blumenhagen:2005zg}
\bibitem{Blumenhagen:2005zg}
  R.~Blumenhagen, G.~Honecker and T.~Weigand,
  %``Non-abelian brane worlds: The heterotic string story,''
  JHEP {\bf 0510}, 086 (2005)
  [arXiv:hep-th/0510049].
  %%CITATION = JHEPA,0510,086;%%



%\cite{Blumenhagen:2005zh}
\bibitem{Blumenhagen:2005zh}
  R.~Blumenhagen, G.~Honecker and T.~Weigand,
  %``Non-abelian brane worlds: The open string story,''
  arXiv:hep-th/0510050.
  %%CITATION = HEP-TH/0510050;%%


%\cite{Katz:2002gh}
\bibitem{Katz:2002gh}
  S.~H.~Katz and E.~Sharpe,
  %``D-branes, open string vertex operators, and Ext groups,''
  Adv.\ Theor.\ Math.\ Phys.\  {\bf 6}, 979 (2003)
  [arXiv:hep-th/0208104].
  %%CITATION = 00203,6,979;%%


%\cite{Witten:1999eg}
\bibitem{Witten:1999eg}
  E.~Witten,
  %``World-sheet corrections via D-instantons,''
  JHEP {\bf 0002}, 030 (2000)
  [arXiv:hep-th/9907041].
  %%CITATION = JHEPA,0002,030;%%














%\cite{Blumenhagen:2006wj}
\bibitem{Blumenhagen:2006wj}
  R.~Blumenhagen, S.~Moster, R.~Reinbacher and T.~Weigand,
  %``Massless spectra of three generation U(N) heterotic string vacua,''
  JHEP {\bf 0705}, 041 (2007)
  [arXiv:hep-th/0612039].
  %%CITATION = JHEPA,0705,041;%%



%\cite{Donagi:1999gc}
\bibitem{Donagi:1999gc}
  R.~Donagi, A.~Lukas, B.~A.~Ovrut and D.~Waldram,
  %``Holomorphic vector bundles and non-perturbative vacua in M-theory,''
  JHEP {\bf 9906}, 034 (1999)
  [arXiv:hep-th/9901009].
  %%CITATION = JHEPA,9906,034;%%

%\cite{Donagi:1996yf}
\bibitem{Donagi:1996yf}
  R.~Donagi, A.~Grassi and E.~Witten,
  %``A non-perturbative superpotential with E(8) symmetry,''
  Mod.\ Phys.\ Lett.\  A {\bf 11}, 2199 (1996)
  [arXiv:hep-th/9607091].
  %%CITATION = MPLAE,A11,2199;%%
\bibitem{Aharony:2007db}
  O.~Aharony, S.~Kachru and E.~Silverstein,
  %``Simple Stringy Dynamical SUSY Breaking,''
  arXiv:0708.0493 [hep-th].
  %%CITATION = ARXIV:0708.0493;%%

%\cite{Aganagic:2007py}
\bibitem{Aganagic:2007py}
  M.~Aganagic, C.~Beem and S.~Kachru,
  %``Geometric Transitions and Dynamical SUSY Breaking,''
  arXiv:0709.4277 [hep-th].
  %%CITATION = ARXIV:0709.4277;%%

%\cite{Cvetic:2003yd}
\bibitem{Cvetic:2003yd}
  M.~Cveti\v c, P.~Langacker and J.~Wang,
  %``Dynamical supersymmetry breaking in standard-like models with  
%intersecting
  %D6-branes,''
  Phys.\ Rev.\  D {\bf 68}, 046002 (2003)
  [arXiv:hep-th/0303208].
  %%CITATION = PHRVA,D68,046002;%%

%\cite{Kallosh:2006dv}
\bibitem{Kallosh:2006dv}
  R.~Kallosh and A.~Linde,
  %``O'KKLT,''
  JHEP {\bf 0702} (2007) 002
  [arXiv:hep-th/0611183].
  %%CITATION = JHEPA,0702,002;%%



%\cite{Dudas:2006gr}
\bibitem{Dudas:2006gr}
  E.~Dudas, C.~Papineau and S.~Pokorski,
  %``Moduli stabilization and uplifting with dynamically generated F-terms,''
  JHEP {\bf 0702}, 028 (2007)
  [arXiv:hep-th/0610297].
  %%CITATION = JHEPA,0702,028;%%

%\cite{Abe:2006xp}
\bibitem{Abe:2006xp}
  H.~Abe, T.~Higaki, T.~Kobayashi and Y.~Omura,
  %``Moduli stabilization, F-term uplifting and soft supersymmetry breaking
  %terms,''
  Phys.\ Rev.\  D {\bf 75}, 025019 (2007)
  [arXiv:hep-th/0611024].
  %%CITATION = PHRVA,D75,025019;%%




%\end{references}
\end{thebibliography}
\end{document}